\documentstyle[12pt,a4,cite,epsf,rotate]{article}

\newcommand{\eps}{\varepsilon}
\newcommand{\epe}{\varepsilon'/\varepsilon}
\newcommand{\mt}{m_{\rm t}}
\newcommand{\mc}{m_{\rm c}}
\newcommand{\ms}{m_{\rm s}}
\newcommand{\md}{m_{\rm d}}
\newcommand{\mw}{M_{\rm W}}
\newcommand{\gev}{\, {\rm GeV}}
\newcommand{\mev}{\, {\rm MeV}}
\newcommand{\bsi}{B_6^{(1/2)}}
\newcommand{\bei}{B_8^{(3/2)}}

\newcommand{\Lms}{\Lambda_{\overline{\rm MS}}}
\newcommand{\eqn}[1]{(\ref{#1})}

\newcommand{\IM}{{\rm Im}}

\newcommand{\vub}{|V_{ub}/V_{cb}|}

\begin{document}

% \bibliographystyle{physics}

%%%%%%%%%%%%%%%%%%%%%%%%%%%%%%%%%%%%%%%%%%%%%%%%%%%%%%%%%%%%%%%%%%%%%%%%%
% The titlepage
%%%%%%%%%%%%%%%%%%%%%%%%%%%%%%%%%%%%%%%%%%%%%%%%%%%%%%%%%%%%%%%%%%%%%%%%%
\setcounter{footnote}{1}

\date{\small (August 1996)}

\author{
{\normalsize\bf Andrzej J.~Buras${}^{1,2}$, Matthias Jamin${}^{3}$,
Markus E.~Lautenbacher${}^{1}$} \\
\ \\
{\small\sl ${}^{1}$ Physik Department, Technische Universit\"at M\"unchen,} \\
{\small\sl D-85748 Garching, Germany} \\
{\small\sl ${}^{2}$ Max-Planck-Institut f\"ur Physik
           -- Werner-Heisenberg-Institut,} \\
{\small\sl F\"ohringer Ring 6, D-80805 M\"unchen, Germany} \\
{\small\sl ${}^{3}$ Institut f\"ur Theoretische Physik, Universit\"at
           Heidelberg,} \\
{\small\sl Philosophenweg 16, D-69120 Heidelberg, Germany}
}

\title{
{\small\sf
\rightline{MPI-Ph/96-57}
\rightline{TUM-T31-94/96}
\rightline{HD-THEP-96-23}
\rightline{hep-ph/9608365}
}
\bigskip
\bigskip
{\Large\sf
A 1996 Analysis of the CP Violating Ratio $\epe$\footnote{
Supported by the German Bundesministerium f\"ur Bildung und Forschung
under contract 06 TM 743 and DFG Project Li 519/2-1.}
}
}

\maketitle
\thispagestyle{empty}

\phantom{xxx} \vspace{-5mm}

\begin{abstract}
\noindent
We update our 1993 analysis of the CP violating ratio $\epe$ in view of
the changes in several input parameters, in particular the improved
value of the top quark mass. We also investigate the strange quark mass
$m_s$ dependence of $\epe$ in view of rather low values found in the most
recent lattice calculations. 
 A simple scanning of the input parameters
within one standard deviation gives the ranges:  $-1.2 \cdot 10^{-4}
\le \epe \le 16.0 \cdot 10^{-4}$ and  $0
\le \epe \le 43.0 \cdot 10^{-4}$ for $\ms(\mc)=150\pm20~ \mev$ and
$\ms(\mc)=100\pm20~ \mev$ respectively.  If the experimentally measured
numbers and the theoretical input parameters are used with Gaussian
errors, we find $\epe= ( 3.6\pm 3.4) \cdot 10^{-4}$ 
and  $\epe= ( 10.4\pm 8.3) \cdot 10^{-4}$ respectively. We also give
results for ${\rm Im} V_{ts}^*V_{td}$.
Analyzing the dependence of $\epe$
on various parameters we find that only for  
$\ms(\mc)\le 100\,\mev$ and a conspiration of
other input parameters, values for $\epe$ as large as $(2-4) \cdot 10^{-3}$ 
and consistent with the NA31 result can be obtained.
\end{abstract}

\newpage
\setcounter{page}{1}
\setcounter{footnote}{0}

%%%%%%%%%%%%%%%%%%%%%%%%%%%%%%%%%%%%%%%%%%%%%%%%%%%%%%%%%%%%%%%%%%%%%%%%%%%%%%%
% The main part of the paper
%%%%%%%%%%%%%%%%%%%%%%%%%%%%%%%%%%%%%%%%%%%%%%%%%%%%%%%%%%%%%%%%%%%%%%%%%%%%%%%

The measurement of $\varepsilon'/\varepsilon$ at the $10^{-4}$ level
remains as one of the important targets of contemporary particle
physics. A non-vanishing value of this ratio would give the first
signal for the direct CP violation ruling out the superweak models.
The experimental situation on Re($\varepsilon'/\varepsilon$) is unclear
at present. While the result of NA31 collaboration at CERN with
Re$(\varepsilon'/\varepsilon) = (23 \pm 7)\cdot 10^{-4}$ \cite{barr:93}
clearly indicates direct CP violation, the value of E731 at Fermilab,
Re$(\varepsilon'/\varepsilon) = (7.4 \pm 5.9)\cdot 10^{-4}$
\cite{gibbons:93}, is compatible with superweak theories
\cite{wolfenstein:64} in which $\varepsilon'/\varepsilon = 0$.
 Hopefully, in about two years the experimental situation concerning
$\varepsilon'/\varepsilon$ will be clarified through the improved
measurements by the two collaborations at the $10^{-4}$ level and by
the KLOE experiment at  DA$\Phi$NE.

There is no question about  that the direct CP violation is present in
the standard model. Yet, accidentally it could turn out that it will be
difficult to see it in $K \to \pi\pi$ decays.  Indeed in the standard
model $\varepsilon'/\varepsilon $ is governed by QCD penguins and
electroweak (EW) penguins. In spite of being suppressed by
$\alpha/\alpha_s$ relative to QCD penguin contributions, the
electroweak penguin contributions have to be included because of the
additional enhancement factor ${\rm Re}A_0/{\rm Re}A_2=22$ relative
to QCD penguins. With increasing $\mt$ the EW penguins become
increasingly important \cite{flynn:89,buchallaetal:90}, and entering
$\varepsilon'/\varepsilon$ with the opposite sign to QCD penguins
suppress this ratio for large $\mt$. For $\mt\approx 200\,\gev$ the ratio
can even be zero \cite{buchallaetal:90}.  Because of this strong
cancelation between two dominant contributions and due to uncertainties
related to hadronic matrix elements of the relevant local operators, a
precise prediction of $\varepsilon'/\varepsilon$ is not possible at
present.

In spite of all these difficulties, a considerable progress has been
made in this decade to calculate $\varepsilon'/\varepsilon$. First of
all the complete next-to-leading order (NLO) effective hamiltonians for
$\Delta S=1$ \cite{burasetal:92a,burasetal:92d,ciuchini:93}, $\Delta S=2$
\cite{burasjaminweisz:90,herrlichnierste:93,herrlichnierste:95} and
$\Delta B=2$ \cite{burasjaminweisz:90} are now available so that a
complete NLO analysis of $\varepsilon'/\varepsilon$ including
constraints from the observed indirect CP violation ($\varepsilon_K$)
and the $B^0_d-\bar B^0_d$ mixing ($\Delta m_{B_d}$) is possible. The improved
determination of the $V_{ub}$ and $V_{cb}$ elements of the CKM matrix
\cite{neubert:95,Gibbons}, and in particular the determination of 
the top quark mass
$\mt$ \cite{cdfd0:96} had of course also an important impact on
$\varepsilon'/\varepsilon$. The main remaining theoretical
uncertainties in this ratio are then the poorly known hadronic matrix
elements of the relevant QCD penguin and electroweak penguin operators,
the values of the non-perturbative parameters $B_K$ and $\sqrt{B_B}
F_B$ and as stressed in \cite{burasetal:92d} the values of $\ms$ and
$\Lambda_{\overline{MS}}$.

In 1993 we have presented a detailed NLO analysis of
$\varepsilon'/\varepsilon$ \cite{burasetal:92d} using all information
available at that time. A similar NLO analysis has been made by the
Rome group \cite{ciuchini:92}. In 1995 the latter group has updated their
analysis to predict a very small ratio
 $\varepsilon'/\varepsilon=(3.1\pm 2.5)\cdot 10^{-4}$
 \cite{ciuchini:95}, essentially consistent with the superweak
scenario. An analysis of
 $\varepsilon'/\varepsilon$ with different treatments of hadronic
 matrix elements can be found  in
\cite{heinrichetal:92,bertolinietal:95,paschos:96} and will be briefly
discussed below.

The purpose of the present letter is to update our 1993 analysis and
to confront our new result with the one of the Rome group \cite{ciuchini:95}
and with \cite{heinrichetal:92,bertolinietal:95,paschos:96}.

Let us list the main new ingredients of our present analysis compared
to the previous one:
\begin{itemize}
\item
The value of the top quark mass from CDF and D0 \cite{cdfd0:96},
\item
Updated values for the elements of the CKM matrix such as
$V_{ub}$ and $V_{cb}$ \cite{neubert:95,Gibbons},
\item
New values of the strange quark mass ($\ms$) coming from most recent lattice 
\cite{alltonetal:94,gupta:96, FNAL:96} and QCD sum rule
\cite{jaminmuenz:95,chetyrkinetal:95,narison:95} calculations,
\item
Improved value for $\alpha_s(M_Z)$ \cite{Schmelling}, for which we take
$\alpha_s(M_Z)=0.118\pm0.005$ corresponding to $\Lms^{(4)}=325\pm 80 \mev$,
\item
The inclusion of NLO corrections to the QCD factor $\eta_3$
\cite{herrlichnierste:95} entering the top-charm contribution in the
$\Delta S=2$ effective hamiltonian, the last previously missing
ingredient of a complete NLO analysis of  $\varepsilon'/\varepsilon$, and
\item
Two distinct analyses of theoretical and experimental uncertainties.
\end{itemize}

Let us first  recall the basic formulae used in our new analysis,
referring frequently to our 1993 paper \cite{burasetal:92d} where
further details can be found.  In~\cite{burasetal:92d} we have analyzed
$\epe$ in the Standard Model including leading and next-to-leading
logarithmic contributions to the Wilson coefficient functions of the
relevant local operators \cite{burasetal:92a,burasetal:92d,ciuchini:93}.
Imposing the constraints from the $CP$
conserving $K \rightarrow \pi \pi$ data on the hadronic matrix elements
of these operators we have given numerical results for $\epe$ as a
function of $\Lms$, $\mt$, $\ms$ and two non-perturbative parameters
$\bsi$ and $\bei$ which cannot be fixed by the $CP$ conserving data at
present. These two parameters are defined by
\begin{equation}
\langle Q_6(\mc) \rangle_0 \equiv \bsi \, \langle Q_6(\mc)
\rangle_0^{\rm (vac)}
\qquad
\langle Q_8(\mc) \rangle_2 \equiv \bei \, \langle Q_8(\mc)
\rangle_2^{\rm (vac)} \, ,
\label{eq:1}
\end{equation}
where
\begin{equation}
Q_{6} = \left( \bar s_{\alpha} \, d_{\beta}  \right)_{\rm V-A}
   \sum_{q} \left( \bar q_{\beta} \,  q_{\alpha} \right)_{\rm V+A}
\qquad
Q_{8} = \frac{3}{2} \left( \bar s_{\alpha} \, d_{\beta} \right)_{\rm
V-A}
         \sum_{q} e_{q} \left( \bar q_{\beta} \,  q_{\alpha}\right)_{\rm V+A}
\label{eq:2}
\end{equation}
are the dominant QCD and electroweak penguin operators, respectively.
The subscripts on the hadronic matrix elements denote the isospin of
the final $\pi\pi$-state.  The label ``vac'' stands for the vacuum
insertion estimate of the hadronic matrix elements in question for
which $\bsi=\bei=1$. The same result is found in the large $N$ limit
\cite{bardeen:87,burasgerard:87}. Also lattice calculations give
similar results $\bsi=1.0 \pm 0.2$ \cite{kilcup:91,sharpe:91} and $\bei
= 1.0 \pm 0.2$ \cite{kilcup:91,sharpe:91,bernardsoni:89,francoetal:89},
$\bei= 0.81(1)$ \cite{gupta2:96}.
We have demonstrated in \cite{burasetal:92d} that in QCD the parameters
$\bsi$ and $\bei$ depend only very weakly on the renormalization scale
$\mu$ when $\mu > 1\gev$ is considered. The $\mu$ dependence for the
matrix elements in \eqn{eq:1} is then given to an excellent accuracy by
$1/\ms^2(\mu)$ with $\ms(\mu)$ denoting the running strange quark
mass. The scale $\mu=\mc$ in \eqn{eq:1} is a convenient choice for
the extraction of matrix elements from the CP conserving data.

At this point it seems appropriate to summarize the present status of 
the value of the strange quark mass. The most recent results of QCD sum
rule (QCDSR) calculations \cite{jaminmuenz:95,chetyrkinetal:95,narison:95}
obtained at $\mu=1~GeV$ correspond to $\ms(\mc)=170\pm20\,\mev$ with
$\mc=1.3\,\gev$. The lattice calculation of \cite{alltonetal:94} finds
$\ms(2\,\gev)=128\pm18\,\mev$ which corresponds to $\ms(\mc)=150\pm20\,\mev$,
in rather good agreement with the QCDSR result. This summer a new lattice
result has been presented by Gupta and Bhattacharya \cite{gupta:96}.
In the quenched approximation they
find $\ms(2\,\gev)=90\pm20\,\mev$ corresponding to $\ms(\mc)=105\pm20\,\mev$.
For $n_f=2$ the value is found to be even lower: $\ms(2\,\gev)=70\pm15\,\mev$
corresponding to $\ms(\mc)=82\pm17\,\mev$. Similar results are expected
to come soon from the lattice group at FNAL\cite{FNAL:96}.
Certainly these results are on the low side of all strange quark mass
determinations.

Now, as an average from quenched lattice QCD, we can take
$\ms(\mc)=130\pm20\,\mev$ and when averaged with the QCD sum rule value,
we arrive at $\ms(\mc)=150\pm20\,\mev$. This will be one of the ranges
for the strange quark mass to be used in our analysis. On the other
hand we cannot exclude at present that the ultimate values for $\ms$
will be as low as found in the most recent lattice calculations
\cite{gupta:96,FNAL:96}. In order to cover this possibility we will
also present results for $\ms(\mc)=100\pm20\,\mev$. In 
table \ref{tab:ms} we
provide the dictionary between the values of $\ms$ normalized at
different scales. To this end we have used the standard renormalization
group formula at two-loop level with $\Lms^{(4)}=325~\mev$.

\begin{table}[thb]
\caption[]{The dictionary between the values of $\ms$ in units of
$\mev$ normalized at different scales with $\mc=1.3~\gev$.
\label{tab:ms}}
\begin{center}
\begin{tabular}{|c|c|c|c|c|c|}\hline
  $\ms(\mc)$& $ ~75$& $100$& $125$ & $150$ &  $175$ \\ \hline
 $\ms(2~\gev)$& $ ~65$& $~86$& $108$ & $129$ &  $151$ \\ \hline
 $\ms(1~\gev)$& $ ~87$& $116$& $144$ & $173$ &  $202$ \\ \hline
 \end{tabular}
\end{center}
\end{table}

It should also be remarked that the decomposition of the relevant hadronic
matrix elements of penguin operators into a product of $B_i$ factors times
$1/m_s^2$ although useful in the $1/N$ approach is unnecessary in a brute
force method like the lattice approach. It is to be expected that the
future lattice calculations will directly give the relevant hadronic 
matrix elements and the issue of $\ms$ in connection with $\epe$ will
effectively disappear.

The details of the calculation of the Wilson coefficient functions as well
 as the determination of the hadronic matrix elements from the CP-conserving 
data can be found in \cite{burasetal:92d} and will not be repeated here. 
We will rather
present an update of the analytic formula for $\epe$ of ref.
\cite{buraslauten:93} which to a very good accuracy represents our
numerical analysis. This analytic formula exhibits the
dependence of $\epe$ on $\mt$, $\ms$, $\Lms^{(4)}$, $\bsi$ and $\bei$.
It is given as follows:

\begin{equation}
\epe = {\rm Im}\lambda_t \, F(x_t) \, ,
\label{eq:3}
\end{equation}
where
\begin{equation}
F(x_t) =
P_0 + P_X \, X_0(x_t) + P_Y \, Y_0(x_t) + P_Z \, Z_0(x_t) 
+ P_E \, E_0(x_t) \, ,
\label{eq:3b}
\end{equation}
and
\begin{equation}
{\rm Im}\lambda_t = {\rm Im} V_{ts}^*V_{td} = |V_{\rm ub}| \, 
|V_{\rm cb}| \, \sin \delta = \eta \, \lambda^5 \, A^2
\label{eq:4}
\end{equation}
in the standard parameterization of the CKM matrix
\cite{particledata:92} and in the Wolfenstein parameterization
\cite{wolfenstein:83}, respectively. 

The $\mt$-dependent functions in (\ref{eq:3b}) are given by
\begin{eqnarray}
X_0(x_t) &=& \frac{x_t}{8} \,
\left[ \frac{x_t+2}{x_t-1} + \frac{3 x_t -6}{(x_t-1)^2} \ln x_t \right]
\nonumber \\
Y_0(x_t) &=& \frac{x_t}{8} \,
\left[ \frac{x_t-4}{x_t-1} + \frac{3 x_t}{(x_t-1)^2} \ln x_t \right] 
\label{eq:5} \\
Z_0(x_t) &=& 
-\frac{1}{9} \ln x_t +
\frac{18 x_t^4 - 163 x_t^3 + 259 x_t^2 - 108 x_t}{144 (x_t-1)^3} +
\nonumber \\
 & &
\frac{32 x_t^4 - 38  x_t^3 - 15  x_t^2 +  18 x_t}{72  (x_t-1)^4} \ln x_t
\nonumber \\
E_0(x_t) &=& 
-\frac{2}{3} \ln x_t + 
\frac{x_t^2 (15 - 16 x_t + 4 x_t^2)}{6 (1-x_t)^4} \ln x_t +
\frac{x_t (18 - 11 x_t - x_t^2)}{12 (1-x_t)^3} \nonumber
\end{eqnarray}
with $x_t = \mt^2/\mw^2$. In the range $150\gev \le \mt \le 200\gev$
these functions can be approximated to better than 1\% accuracy by the
following expressions \cite{buchallaetal:95}
\begin{eqnarray}
X_0(x_t) = 0.660 \, x_t^{0.575} &\qquad&
Y_0(x_t) = 0.315 \, x_t^{0.78} \label{eq:6} \\
Z_0(x_t) = 0.175 \, x_t^{0.93} &\qquad&
E_0(x_t) = 0.564 \, x_t^{-0.51} \nonumber \, .
\end{eqnarray}

The coefficients $P_i$ are given in terms of $B_6^{(1/2)} \equiv
B_6^{(1/2)}(\mc)$, $B_8^{(3/2)} \equiv B_8^{(3/2)}(\mc)$ and $\ms(\mc)$
as follows
\begin{equation}
P_i = r_i^{(0)} + \left[ \frac{158\mev}{\ms(\mc)+\md(\mc)} \right]^2
\left(r_i^{(6)} B_6^{(1/2)} + r_i^{(8)} B_8^{(3/2)} \right) \, .
\label{eq:pbePi}
\end{equation}
The $P_i$ are renormalization scale and scheme independent. They depend
however on $\Lms$. In table~\ref{tab:pbendr} we give the numerical
values of $r_i^{(0)}$, $r_i^{(6)}$ and $r_i^{(8)}$ for different values
of $\Lms^{(4)}$ at $\mu=\mc$ in the NDR renormalization scheme. 
The
coefficients $r_i^{(0)}$, $r_i^{(6)}$ and $r_i^{(8)}$ depend only very
weakly on
$\ms(\mc)$ as the dominant $\ms$ dependence has been factored out. The
numbers given in table~\ref{tab:pbendr} correspond to $\ms(\mc)=150\,\mev$.
However, even for $\ms(\mc)\approx100\mev$ the analytic expressions given
here reproduce our numerical calculations of $\epe$ to better than $4\%$.
For different scales $\mu$ the numerical values in the tables change
without modifying the values of the $P_i$'s as it should be. To this
end also $B_6^{(1/2)}$ and $B_8^{(3/2)}$ have to be modified as they
depend albeit weakly on $\mu$.

Concerning the scheme dependence only the $r_0$ coefficients
are scheme dependent at the NLO level. Their values in the HV
scheme are given in the last row of table~\ref{tab:pbendr}.
The coefficients $r_i$, 
$i=X, Y, Z, E$ are on the other hand scheme independent at NLO. 
This is related to the fact that the $\mt$
dependence in $\epe$ enters first at the NLO level and consequently all
coefficients $r_i$ in front of the $\mt$ dependent functions must be
scheme independent. 

Consequently, when changing the renormalization scheme one is only
obliged to change appropriately $B_6^{(1/2)}$ and $B_8^{(3/2)}$ in the
formula for $P_0$ in order to obtain a scheme independence of $\epe$.
In calculating $P_i$ where $i \not= 0$, $B_6^{(1/2)}$ and $B_8^{(3/2)}$
can in fact remain unchanged, because their variation in this part
corresponds to higher order contributions to $\epe$ which would have to
be taken into account in the next order of perturbation theory.

For similar reasons the NLO analysis of $\epe$ is still insensitive to
the precise definition of $\mt$. In view of the fact that the NLO
calculations needed to extract $\IM \lambda_t$ (see below) have been 
done with $\mt=\overline{m}_t(\mt)$ we will also use  this 
definition in calculating $F(x_t)$. The value for 
$\overline{m}_t(\mt)$ corresponding to the average pole mass
$\mt^{pole}=175\pm 6\,\gev$ from  CDF and D0 \cite{cdfd0:96}, 
is $\overline{\mt}(\mt)=167 \pm 6\gev$. 
In what follows $\mt$ stands always for $\overline{m}_t(\mt)$.

\begin{table}[thb]
\caption[]{$\Delta S=1$ PBE coefficients for various $\Lms$ in the NDR scheme.
The last row gives the $r_0$ coefficients in the HV scheme.
\label{tab:pbendr}}
\begin{center}
\begin{tabular}{|c||c|c|c||c|c|c||c|c|c|}
\hline
& \multicolumn{3}{c||}{$\Lms^{(4)}=245\mev$} &
  \multicolumn{3}{c||}{$\Lms^{(4)}=325\mev$} &
  \multicolumn{3}{c| }{$\Lms^{(4)}=405\mev$} \\
\hline
$i$ & $r_i^{(0)}$ & $r_i^{(6)}$ & $r_i^{(8)}$ &
      $r_i^{(0)}$ & $r_i^{(6)}$ & $r_i^{(8)}$ &
      $r_i^{(0)}$ & $r_i^{(6)}$ & $r_i^{(8)}$ \\
\hline
0 &
   --2.674 &   6.537 &   1.111 &
   --2.747 &   8.043 &   0.933 &
   --2.814 &   9.929 &   0.710 \\
$X$ &
    0.541 &   0.011 &       0 &
    0.517 &   0.015 &       0 &
    0.498 &   0.019 &       0 \\
$Y$ &
    0.408 &   0.049 &       0 &
    0.383 &   0.058 &       0 &
    0.361 &   0.068 &       0 \\
$Z$ &
    0.178 &  --0.009 &  --6.468 &
    0.244 &  --0.011 &  --7.402 &
    0.320 &  --0.013 &  --8.525 \\
$E$ &
    0.197 &  --0.790 &   0.278 &
    0.176 &  --0.917 &   0.335 &
    0.154 &  --1.063 &   0.402 \\
\hline
0 &
   --2.658 &   5.818 &   0.839 &
   --2.729 &   6.998 &   0.639 &
   --2.795 &   8.415 &   0.398 \\
\hline
\end{tabular}
\end{center}
\end{table}

The inspection of the table~\ref{tab:pbendr} shows
that the terms involving $r_0^{(6)}$ and $r_Z^{(8)}$ dominate the ratio
$\epe$. The function $Z_0(x_t)$ representing a gauge invariant
combination of $Z^0$- and $\gamma$-penguins grows rapidly with $\mt$
and due to $r_Z^{(8)} < 0$ these contributions suppress $\epe$ strongly
for large $\mt$ \cite{flynn:89,buchallaetal:90}. 

In order to complete our analysis of $\epe$ we need the value of ${\rm
Im}\lambda_t$. To this end we will use the standard expression for
$\eps_K$ describing the indirect CP violation in $K \to \pi\pi$ and the
corresponding expression for $\Delta m_{B_d}$ which describes the
$B^0_d$-$\bar B^0_d$ mixing. Since these expressions are by now well known
we will not repeat them here. They can be found in \cite{buchallaetal:95}.
We list here only the QCD factors $\eta_i\,(i=1,2,3)$ and $\eta_B$
relevant for $\eps_K$ and $x_d$ respectively:
\begin{equation}
\eta_1 = 1.38 
\qquad
\eta_2 = 0.57
\qquad
\eta_3 = 0.47 
\qquad
\eta_B =0.55
\label{eq:etaknum}
\end{equation}
They all include NLO corrections
\cite{burasjaminweisz:90,herrlichnierste:93,herrlichnierste:95}.

The resulting value for ${\rm Im}\lambda_t$ depends on several input
parameters such as $|V_{cb}|$, $|V_{ub}/V_{cb}|$, $\mt$, $B_K$ and
$\sqrt{B_{B_d}} F_{B_d}$.  Here $B_i$ are the non-perturbative
parameters related to the hadronic matrix elements of the $\Delta S=2$
and $\Delta B=2$ operators and $F_{B_d}$ is the $B_d$ meson decay
constant.  Values and errors of these input parameters, used in the
present analysis, are collected in table~\ref{tab:inputparams} together
with the experimental values for $\eps_K$ \cite{particledata:92}
 and $\Delta m_{B_d}$ \cite{Gibbons}. Except for $\mt$, $\eps_K$,
$\Delta m_{B_d}$ and $\Lms^{(4)}$ all input quantities are the same as in 
the review
\cite{buchallaetal:95} where further details on the chosen ranges with
relevant references can be found. See also \cite{Flynn}. For 
$\tau(B_d)=1.6~ps$ the value for $\Delta m_{B_d}$ 
in table~\ref{tab:inputparams} corresponds to the mixing parameter
$x_d=0.74\pm 0.03$ 

\begin{table}[thb]
\caption[]{Collection of input parameters.\label{tab:inputparams}}
\begin{center}
\begin{tabular}{|c|c|c|}
\hline
{\bf Quantity} & {\bf Central} & {\bf Error} \\
\hline
$|V_{cb}|$ & 0.040 & $\pm 0.003$ \\
$|V_{ub}/V_{cb}|$ & 0.080 & $\pm 0.020$ \\
$B_K$ & 0.75 & $\pm 0.15$ \\
$\sqrt{B_d} F_{B_{d}}$ & $200\mev$ & $\pm 40\mev$ \\
$\Lms^{(4)}$ & $325 \mev$& $ \pm 80 \mev$ \\ 
$\mt$ & $167\gev$ & $\pm 6\gev$ \\
$\Delta m_{B_d}$ & $0.464~ps^{-1}$ & $\pm 0.018~ps^{-1}$ \\
$\eps_K$ & $2.280 \cdot 10^{-3}$ & $\pm 0.013 \cdot 10^{-3}$ \\
\hline
\end{tabular}
\end{center}
\end{table}

In what follows we will present two types of the numerical analyses of
${\rm Im}\lambda_t$ and $\epe$:

\begin{itemize}
\item
Method 1: Both the experimentally measured numbers and the theoretical input
parameters are scanned independently within the errors given in
table~\ref{tab:inputparams}.
\item
Method 2: The experimentally measured numbers and the theoretical input 
parameters are used with Gaussian errors.
\end{itemize}

\begin{figure}[hbt]
\vspace{0.10in}
\centerline{
\epsfysize=4.5in
\rotate[r]{
\epsffile{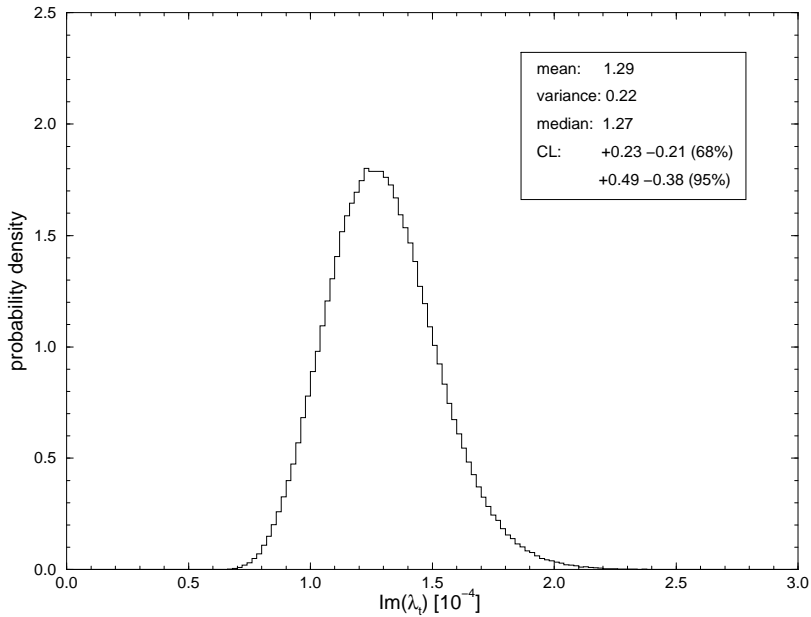}
}}
\vspace{0.08in}
\caption[]{
Probability density distribution for ${\rm Im}\lambda_t$ using input
parameters as given in the text.
\label{fig:mcimlt}}
\end{figure}

\begin{figure}[hbt]
\vspace{0.10in}
\centerline{
\epsfysize=4.5in
\rotate[r]{
\epsffile{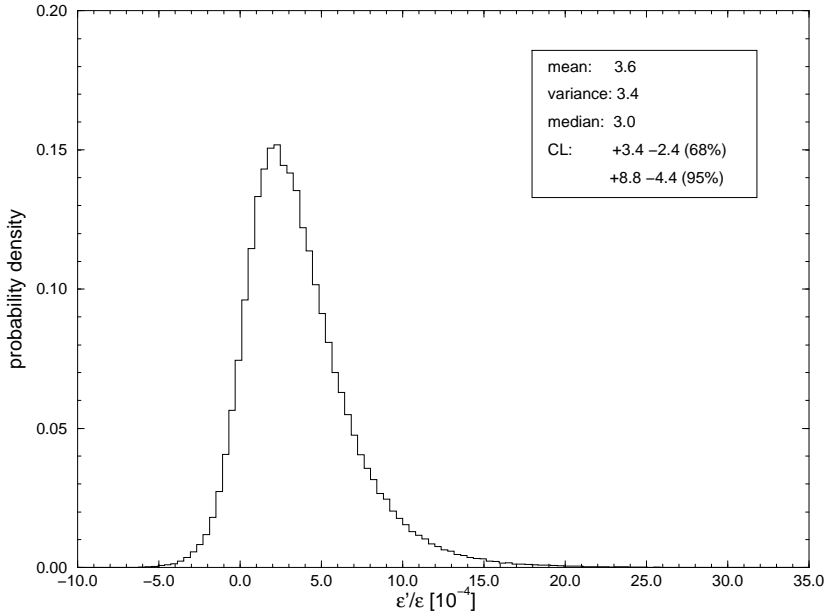}
}}
\vspace{0.08in}
\caption[]{
Probability density distribution for $\epe$ for $\ms(\mc)=150\pm20\,\mev$
using other input parameters as given in the text.
\label{fig:mcepe}}
\end{figure}

The first method is the one used in our 1993 paper as well as in
\cite{buchallaetal:95}. 
The second method is similar to the one used by the Rome group 
\cite{ciuchini:95}
except that these authors assume  a flat distribution (with a
width of $2 \sigma$) for the theoretical quantities. 
Whereas the first method is even more conservative than adding all
errors for the various input parameters linearly, the second method
yields a similar error estimate as if all errors would have been
added in quadrature. Thus the second method should be considered
as reflecting a lower bound on the combined error with the true
uncertainty lying somewhere in between the two methods.

In our new analysis let us first concentrate on the case
$\ms(\mc)=150\pm20\,\mev$.
Using the first method and the parameters in table~\ref{tab:inputparams} 
we find :
\begin{equation}
0.86 \cdot 10^{-4} \le {\rm Im}\lambda_t \le 1.71 \cdot 10^{-4}
\label{eq:imnew}
\end{equation}
\begin{equation}
-1.2 \cdot 10^{-4} \le \epe \le 16.0 \cdot 10^{-4}
\label{eq:eperangenew}
\end{equation}
These ranges are similar to the ones found in \cite{buchallaetal:95}
where slightly larger errors for $m_t$ and $\Delta m_{B_d}$ have been used.

Using next the second method we find the distributions of values for
${\rm Im}\lambda_t$ and $\epe$ in figs.~\ref{fig:mcimlt} and
\ref{fig:mcepe}, respectively. 
From the distributions in figs.~\ref{fig:mcimlt} and \ref{fig:mcepe}
we deduce the following results:
\begin{equation}
 {\rm Im}\lambda_t =( 1.29 \pm 0.22) \cdot 10^{-4}
\label{eq:imfinal}
\end{equation}
\begin{equation}
\epe= ( 3.6\pm 3.4) \cdot 10^{-4}
\label{eq:eperangefinal}
\end{equation} 
In addition in the figures we have also given central values according
to the median prescription and the corresponding 1 and $2\,\sigma$
confidence intervals. Within the errors, they agree with mean and
variance. We observe that the distribution of the values of $\epe$
is asymmetric with a longer tail towards larger values. However
at $95\%$ C.L. small negative values cannot be excluded.

The above results for $\epe$ apply to the NDR scheme.
$\epe$ is generally lower in the HV scheme if the same values for
$B_6^{(1/2)}$ and $B_8^{(3/2)}$ are used in both schemes. In view of the
fact that the differences between NDR and HV schemes are smaller than
the uncertainties in $B_6^{(1/2)}$ and $B_8^{(3/2)}$ we think it is
sufficient to present only the results in the NDR scheme here. The
results in the HV scheme can be found in
\cite{burasetal:92d,ciuchini:95}.

In spite of some differences in the treatment of hadronic matrix
elements our results for $\epe$, with $\ms(\mc)=150\pm 20 \mev$,
 using the second method agree well with
the results of the Rome group \cite{ciuchini:95}.
On the other hand the range in (\ref{eq:eperangenew}) shows that for
particular choices of the input parameters, values for $\epe$ as high as
$15\cdot 10^{-4}$ cannot be excluded at present. Such high values are
found if simultaneously  $\vub=0.10$, $B_6=1.2$, $B_8=0.8$, $B_K=0.6$,
$\ms(\mc)=130 MeV$, $\Lms^{(4)}=405\mev$ and low values of $m_t$ still
consistent with $\varepsilon_K$ and the observed $B^0_d-\bar B^0_d$ mixing
are chosen. It is however evident from  figure \ref{fig:mcepe} that such 
high values of $\epe$ and generally values above $10^{-3}$ 
are very improbable.

In \cite{bertolinietal:95} the hadronic matrix elements relevant for
$\epe$ have been calculated within the chiral quark model. Using the
first method the authors find a rather large range $-50 \cdot 10^{-4}
\le \epe \le 14 \cdot 10^{-4}$.  In particular they find in contrast to
\cite{burasetal:92d,ciuchini:95,buchallaetal:95} and the
present analysis that negative values for $\epe$ as large as $-5\cdot
10^{-3}$ are possible. This is related to the fact that in the chiral
quark model $B_6$ and $B_8$ can for certain model parameters deviate
considerably from unity and generally $B_8>B_6$. The Dortmund group
\cite{heinrichetal:92} advocates on the other hand $B_6>B_8$ to find
$\epe=(9.9\pm 4.1)\cdot 10^{-4}$ for $\ms(1\gev)=175\,\mev$ \cite{paschos:96}.
From the point of view of the present analysis and the results of the
Rome group such high values of $\epe$ for $\ms(\mc)={\cal O}(150\,\mev)$
are rather improbable within the standard model.

\begin{figure}[hbt]
\vspace{0.10in}
\centerline{
\epsfysize=4.5in
\rotate[r]{
\epsffile{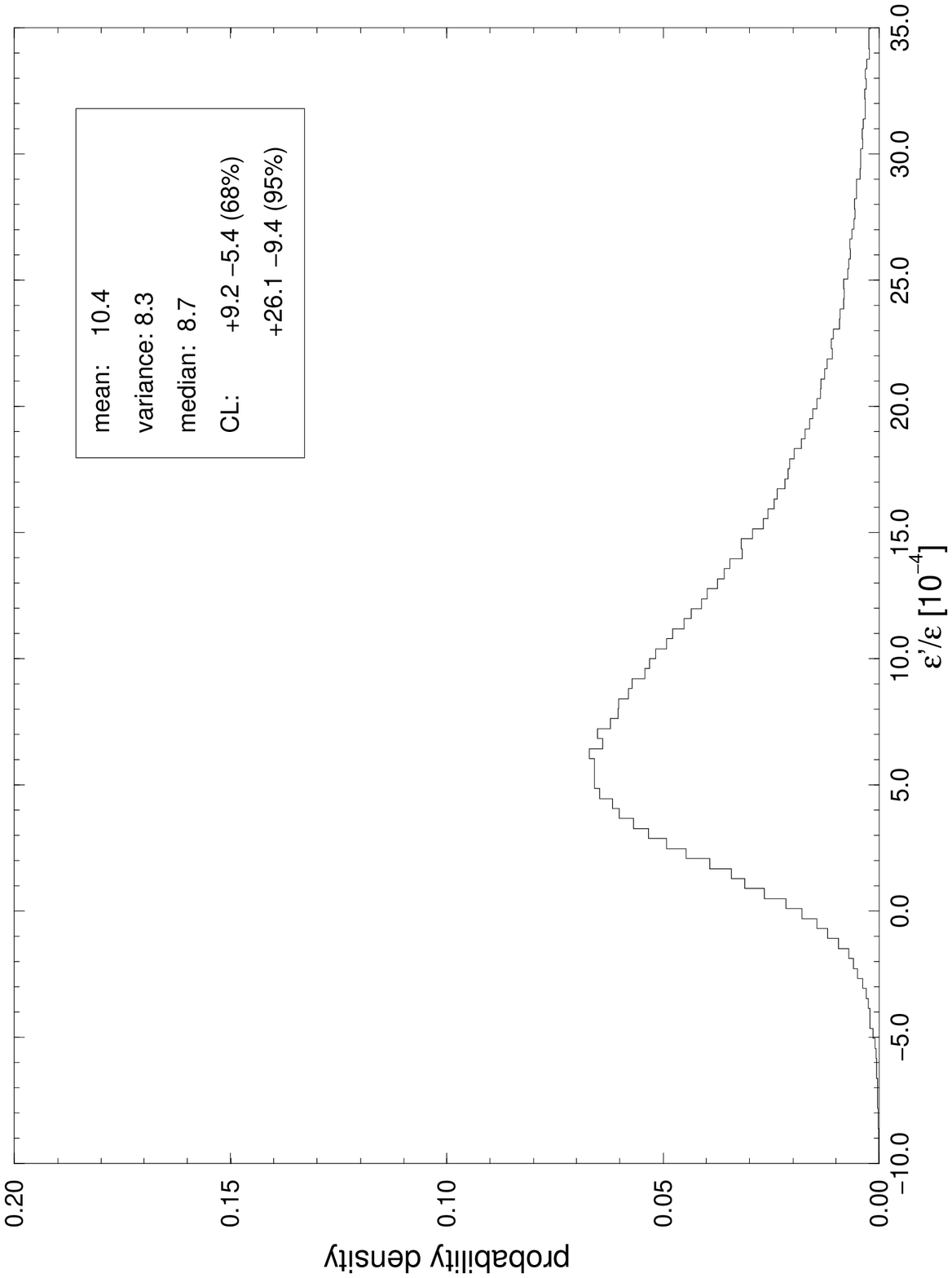}
}}
\vspace{0.08in}
\caption[]{
Probability density distribution for $\epe$ for $\ms(\mc)=100\pm20\,\mev$
using other input parameters as given in the text.
\label{fig:mcepe100}}
\end{figure}

The situation with $\epe$ in the standard model may however change
considerably if the value for $\ms$ is as low as found in 
\cite{gupta:96,FNAL:96}.
Repeating our analysis for $\ms(\mc)=100\pm20\,\mev$ we find
\begin{equation}
0 \le \epe \le 43.0 \cdot 10^{-4}
\label{eq:eperangenew2}
\end{equation}
and
\begin{equation}
\epe= ( 10.4\pm 8.3) \cdot 10^{-4}
\label{eq:eperangefinal2}
\end{equation} 
in place of (\ref{eq:eperangenew}) and
(\ref{eq:eperangefinal}) respectively. The corresponding distribution
is shown in fig.~\ref{fig:mcepe100}. We observe that the resulting 
distribution of
the values of $\epe$ is in this case rather asymmetric with a very 
long tail towards substantial positive values. Moreover,
negative values of $\epe $ are found to be very unlikely.

In addition, it is of interest to investigate for which values of the
input parameters the standard model can reproduce the results of NA31
and E731 collaborations. In table \ref{tab:31731} we present the
values of $\epe$ for five choices of $\ms(\mc)$ and for selective
sets of other input parameters keeping $V_{cb}=0.040$, $\mt=167\,\gev$
and $B_K=0.75$ fixed. The values of $\epe$ given in this table
correspond to $\sin\delta$ in the first quadrant. The results
for the second quadrant turn out to be somewhat lower.
We observe that the decrease of $\ms$ for $\ms(\mc)\geq 100\mev$
alone is insufficient to bring the standard model to agree with
the  values obtained by the NA31 collaboration. For central values
of other parameters and  $\ms(\mc)={\cal O}(100\,\mev)$ the
standard model prediction is rather in the ball park of the E731 result.
However for $B_6>B_8$, sufficiently large values of
$|V_{ub}/V_{cb}|$ and $\Lms$ and small values of $\ms(\mc)$, the values
of $\epe$ in the standard model can be as large as $(2-4)\cdot 10^{-3}$
and consistent with the NA31 result. 

\begin{table}[thb]
\caption[]{ Values of $\epe$ in units of $10^{-4}$ 
for specific values of various input parameters at $\mt=167\,\gev$, 
$V_{cb}=0.040$ and $B_K=0.75$.
\label{tab:31731}}
\begin{center}
\begin{tabular}{|c|c|c|c|c||c|}\hline
  $|V_{ub}/V_{cb}|$& $\Lms^{(4)}\;[\mev]$& $B_6$& $B_8$ & $\ms(\mc)\,[\mev]$ &
 $\epe$ \\ \hline
       &       &      &     & $~75$ & $16.8$ \\
       &       &      &     & $100$ & $~9.1$ \\
$0.08$ & $325$ & $1.0$&$1.0$& $125$ & $~5.3$ \\
       &       &      &     & $150$ & $~3.2$ \\
       &       &      &     & $175$ & $~1.8$ \\ \hline\hline
       &       &      &     & $~75$ & $27.8$ \\
       &       &      &     & $100$ & $15.6$ \\
$0.08$ & $325$ & $1.2$&$0.8$& $125$ & $~9.6$ \\
       &       &      &     & $150$ & $~6.2$ \\
       &       &      &     & $175$ & $~4.1$ \\ \hline\hline
       &       &      &     & $~75$ & $39.8$ \\
       &       &      &     & $100$ & $22.5$ \\
$0.10$ & $405$ & $1.2$&$0.8$& $125$ & $14.0$ \\
       &       &      &     & $150$ & $~9.2$ \\
       &       &      &     & $175$ & $~6.2$ \\ \hline
\end{tabular}
\end{center}
\end{table}

To summarize, we have presented a new analysis of $\epe$, showing
in particular the dependence of this important ratio on various
input parameters, in particular the value of $\ms$. Our results 
are summarized in 
(\ref{eq:eperangenew}) and (\ref{eq:eperangefinal}) for the case
$\ms(\mc)=150\pm20\,\mev$ and in 
(\ref{eq:eperangenew2}) and (\ref{eq:eperangefinal2}) for the case
$\ms(\mc)=100\pm20\,\mev$. 
Furthermore, table \ref{tab:31731} gives values of $\epe$ for particular
sets of input parameters. The improved results for ${\rm Im}\lambda_t$
are given in (\ref{eq:imnew}) and (\ref{eq:imfinal}).

It is clear that the fate of $\epe$ in the standard model after the
improved measurement of $\mt$, depends sensitively on the values of
$|V_{ub}/V_{cb}|$, $\Lms$ and in particular on $B_6$, $B_8$ and $\ms$.
For $\ms(\mc)={\cal O}(150\,\mev)$ $\epe$ is generally below $10^{-3}$
in agreement with E731 with central values in the ball park of
a few $10^{-4}$ as found in \cite{ciuchini:95, buchallaetal:95} and here.
However, if the low values of $\ms(\mc)={\cal O}(100\,\mev)$ found in 
\cite{gupta:96,FNAL:96}
are confirmed by other groups in the future, a conspiration of
other parameters may give values as large as $(2-4)\cdot 10^{-3}$ in the
ball park of the NA31 result. 

Let us hope that the future experimental and theoretical results will
be sufficiently accurate to be able to see whether $\epe\not=0$ and
whether the standard model agrees with the data. In any case the
coming years should be very exciting. 

%%%%%%%%%%%%%%%%%%%%%%%%%%%%%%%%%%%%%%%%%%%%%%%%%%%%%%%%%%%%%%%%%%%%%%%%%
% References
%%%%%%%%%%%%%%%%%%%%%%%%%%%%%%%%%%%%%%%%%%%%%%%%%%%%%%%%%%%%%%%%%%%%%%%%%
\clearpage
\newpage
{\small
% \bibliography{../share/2lref}

}

\end{document}